\documentclass[11pt]{article}
\usepackage{moriond,epsfig}

\def\MyPhoto{\pdfimage width 35mm {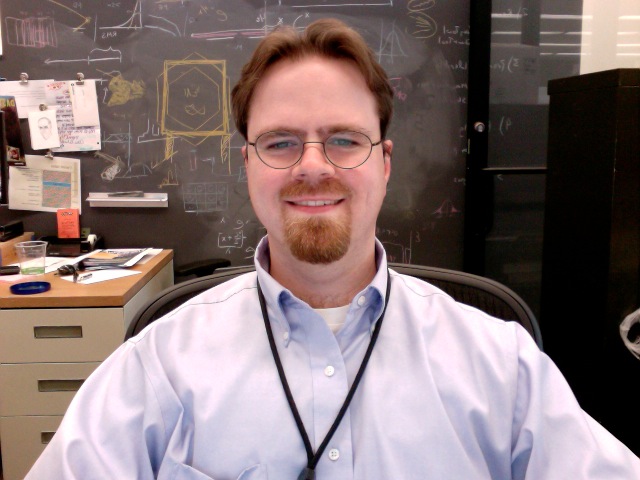}}

\def\be{\begin{equation}}
\def\ee{\end{equation}}
\def\bea{\begin{eqnarray}}
\def\eea{\end{eqnarray}}

\begin{document}

\newcommand{\mnv}{\mathrm{MINER}{\nu}\mathrm{A}}

\vspace*{4cm}
\title{Status Update for the $\mnv$ Experiment}

\author{ Gabriel Perdue \footnote{On Behalf of the $\mnv$ Collaboration.}}

\address{Fermilab, PO Box 500, MS 220 \\
Batavia, IL 60510, USA}

\maketitle\abstracts{
$\mnv$ (Main INjEctoR $\nu$-A) is a few-GeV neutrino cross section experiment that began taking data in the FNAL NuMI beam-line in the fall of 2009.  MINERvA employs a fine-grained detector capable of complete kinematic characterization of neutrino interactions.  The detector consists of an approximately 6.5 ton active target region composed of plastic scintillator with additional carbon, iron, and lead targets upstream of the active region.  The experiment will provide important inputs for neutrino oscillation searches and a pure weak probe of nuclear structure.  Here we offer a set of initial kinematic distributions of interest and provide a general status update.}

\section{Brief Introduction to the $\mnv$ Experiment}\label{sec:intro}
$\mnv$ is a dedicated on-axis neutrino-nucleus scattering experiment running at Fermilab in the NuMI (Neutrinos at the Main Injector) beamline. The primary motivation of $\mnv$ is to accurately measure scattering cross-sections and event kinematics in exclusive and inclusive final states. By including a variety of high and low atomic number (A) targets in the same detector and beam, $\mnv$ will contribute to untangling nuclear effects and determining nuclear parton distribution functions (PDF's).

\subsection{The Era of Precision Neutrino Oscillation Experiments}\label{subsec:precneut}
$\mnv$ results will be important for present and future neutrino oscillation experiments, where cross-sections, final state details, and nuclear effects are all important in calculating incoming neutrino energy and in separating backgrounds from the oscillation signal. Recall that oscillation probability depends on $E_{\nu}$, the neutrino energy. For example, in a $\nu_{\mu}$ disappearance experiment, the two-flavor disappearance relation is show in Eq. \ref{eq:numudisp}:
\begin{equation}
P\left( \nu_{\mu} \rightarrow \nu_{\mu} \right) = 1 - \sin^2 \left( 2\theta_{23} \right) \sin^2 \left( \frac{1.27 \Delta m_{23}^2 (eV^2) L(km)}{E_{\nu}(GeV)} \right)
\label{eq:numudisp}
\end{equation}

However, experiments measure the visible energy of the interaction. Visible energy is a function of flux, cross-section, and detector response. 
Because the neutrino interacts in dense nuclear matter, final state interactions (FSI) play a significant role in the observed final state particles.
Near to Far-Detector ratios cannot handle all of the associated uncertainties because the Near / Far energy spectra are different due to beam, oscillation, matter, and possibly nuclear effects.

\subsection{Nuclear Effects}\label{subsec:nuceff}
As weak force only probes of the nucleus, neutrinos are complementary to charged lepton scattering measurements. There are many quantities of interest with large uncertainties: axial form factors as a function A and momentum transfer ($Q^2$), quark-hadron duality, $x$-dependent nuclear effects, etc. Additionally, $\mnv$ will study nuclear effects in order to understand how interaction probabilities with heavy nuclei differ from those with free nucleons and how to characterize FSI.

\section{The $\mnv$ Detector}\label{sec:detector}
$\mnv$ is a horizontal stack of roughly identical modules weighing on average about two tons each. Modules contain an inner detector (ID) region composed of triangular plastic scintillator strips and an outer detector (OD) steel frame and support structure also instrumented with plastic scintillator bars. Most modules feature an ID composed of two planes of scintillator, but some in the targets and calorimetric regions of the detector give up one or both scintillator planes for target or absorber materials. 
The total nuclear target masses installed as of the end of Winter 2011 are listed in Table \ref{tab:targmass}.

\begin{table}[t]
\caption{$\mnv$ Nuclear Target Masses in the Winter 2011 Run. Note the tracker mass includes the full longitudinal span. For most analyses, the effective mass will be closer to five tons. \label{tab:targmass}}
\vspace{0.4cm}
\begin{center}
\begin{tabular}{|c|c||}
\hline
Target & Fiducial Mass (90 cm radius cut) \\
\hline
Scintillator Tracker (CH) & 6.43 tons \\
\hline
Carbon (Graphite) & 0.17 tons \\
\hline
Iron & 0.97 tons \\
\hline
Lead & 0.98 tons \\
\hline
\end{tabular}
\end{center}
\end{table}

\begin{figure}
\centering
\psfig{figure=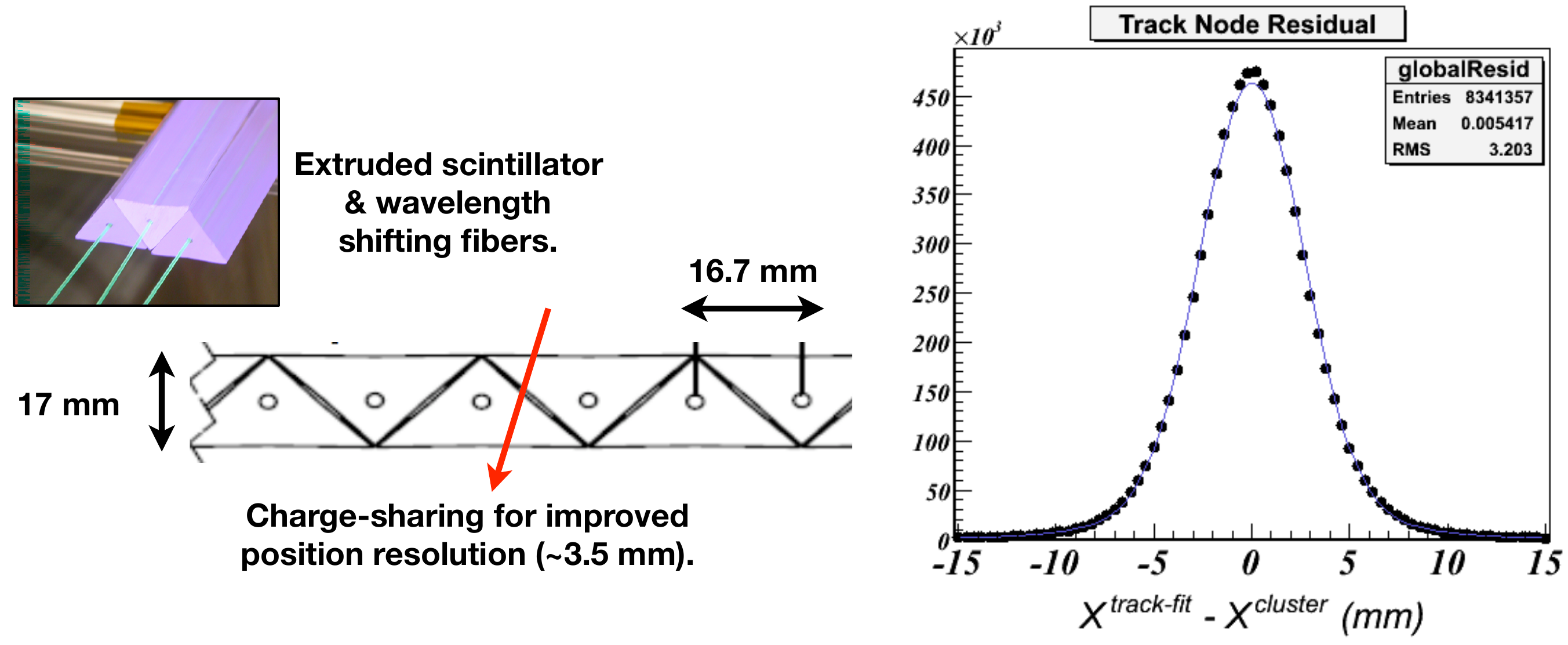,height=2.0in}
\caption{Shown here is an illustration of the triangular scintillator strip arrangement. Groups of 127 strips are bundled into ``planes.'' Also shown are tracking residuals between a fitted position along a track the charge-weighted hit in that plane for a sample of through-going muons. 
\label{fig:triangstrips}}
\end{figure}

\begin{figure}
\centering
\psfig{figure=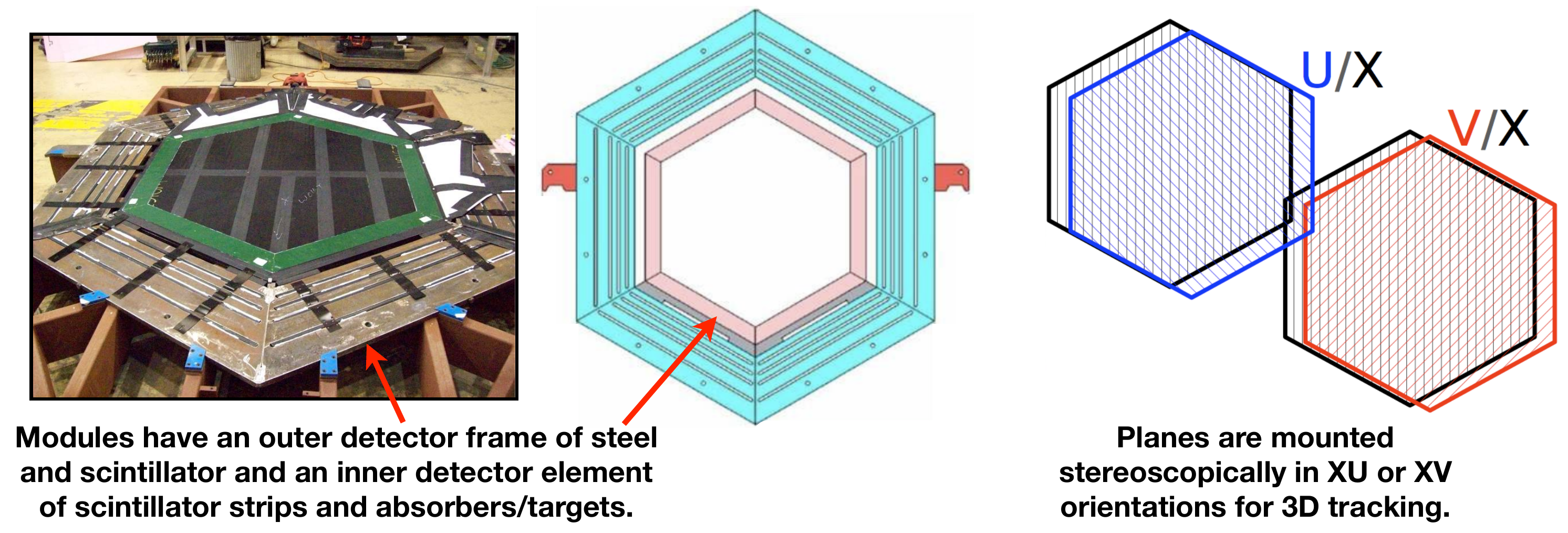,height=1.8in}
\caption{Plane structure illustrated in a photograph from module construction and an engineering diagram. There are three basic orientations for strips in scintillator planes: ``U'' , where the strips are oriented perpendicular to $-\left(60^0\right)$,  ``X'' (strips oriented vertically), and  ``V'', where the strips are oriented perpendicular to $\left(60^0\right)$.
\label{fig:planestruct}}
\end{figure}

See Fig. \ref{fig:triangstrips} for an illustration of the triangular strips used in the ID. We utilize charge sharing to improve position resolution and are able to achieve tracking residuals of just over 3 mm. Fig. \ref{fig:planestruct} shows the arrangement of planes, along with a photograph from module construction. 

\begin{figure}
\centering
\psfig{figure=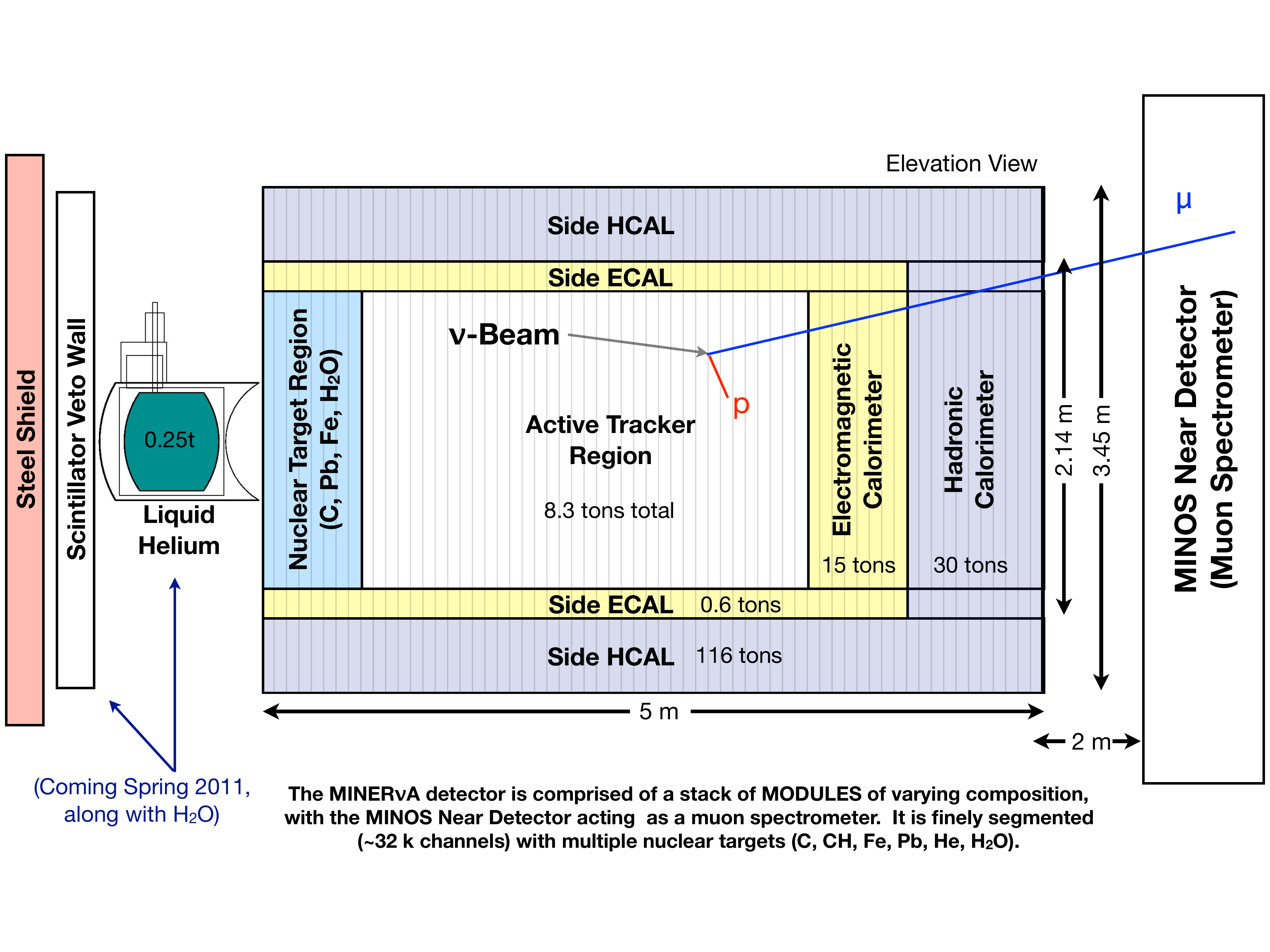,height=3.2in}
\caption{Schematic of the $\mnv$ detector as of winter 2011. The figure is only roughly to scale.
\label{fig:detect01}}
\end{figure}

Planes are composed of strips oriented along the X, U, or V axes. The typical arrangement is a module containing two planes, one with strips in the U or V direction and one with strips in the X direction. 
The stack alternates through the detector, UX, VX, UX, etc. Combination of the X, U, and V views allows three-dimensional reconstruction. Figure \ref{fig:detect01} shows a schematic of the detector layout.

\section{Data Collection}\label{sec:datcoll}
\begin{figure}
\centering
\psfig{figure=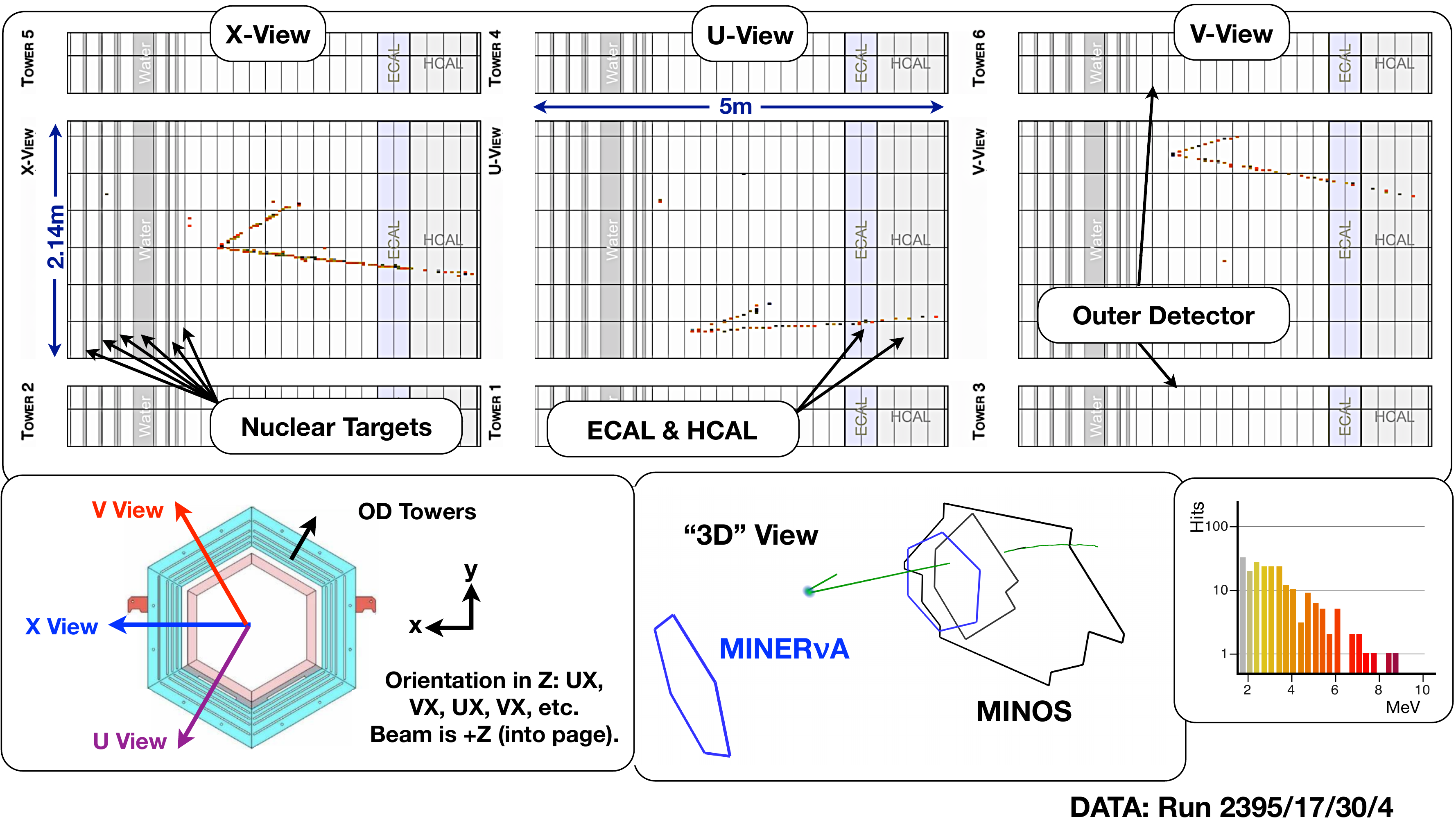,height=2.0in}
\caption{A charged-current interaction candidate event display from data. 
\label{fig:eventdisplay}}
\end{figure}

MINERvA began taking neutrino data in March, 2010, with the NuMI beam line in the forward horn current (FHC) mode, focusing $\pi^{+}$ mesons (``neutrino mode'').  Reverse horn current (RHC) mode data, focusing $\pi^{-}$ mesons (``anti-neutrinos mode''), was taken prior to March, 2010. 
Table \ref{tab:rates} shows raw MC event generator estimates for the event rates in these data samples. Our MC event generator is GENIE 2.6.2 ~\cite{gn}.
Figure \ref{fig:eventdisplay} shows a typical charged-current (CC) neutrino event candidate from our FHC data set. 

\begin{table}[t]
\caption{Charged-current inclusive event rates in the current data sample. (GENIE 2.6.2 Generator raw events, not acceptance corrected.) \label{tab:rates}}
\vspace{0.4cm}
\begin{center}
\begin{tabular}{|c|c|c|}
\hline
Material & $1.2 \times 10^{20}$ P.O.T. LE $\nu$ Mode & $1.2 \times 10^{20}$ P.O.T. LE $\bar{\nu}$ Mode \\
\hline
Carbon Target & 10,800 & 3,400 \\
\hline
Iron Target & 64,500 & 19,200 \\
\hline
Lead Target & 68,400 & 10,800 \\
\hline
Scintillator (CH) Tracker & 409,000 & 134,000 \\
\hline
\end{tabular}
\end{center}
\end{table}

\section{Flux Estimate}\label{sec:flux}
$\mnv$ uses the FNAL NuMI beam-line. One of the key features of the beam-line is the ability to move the target relative to the meson focusing horns and change the current in the horns. 
This allows experiments to tune the energy spectra of the beam. 

\begin{figure}
\centering
\psfig{figure=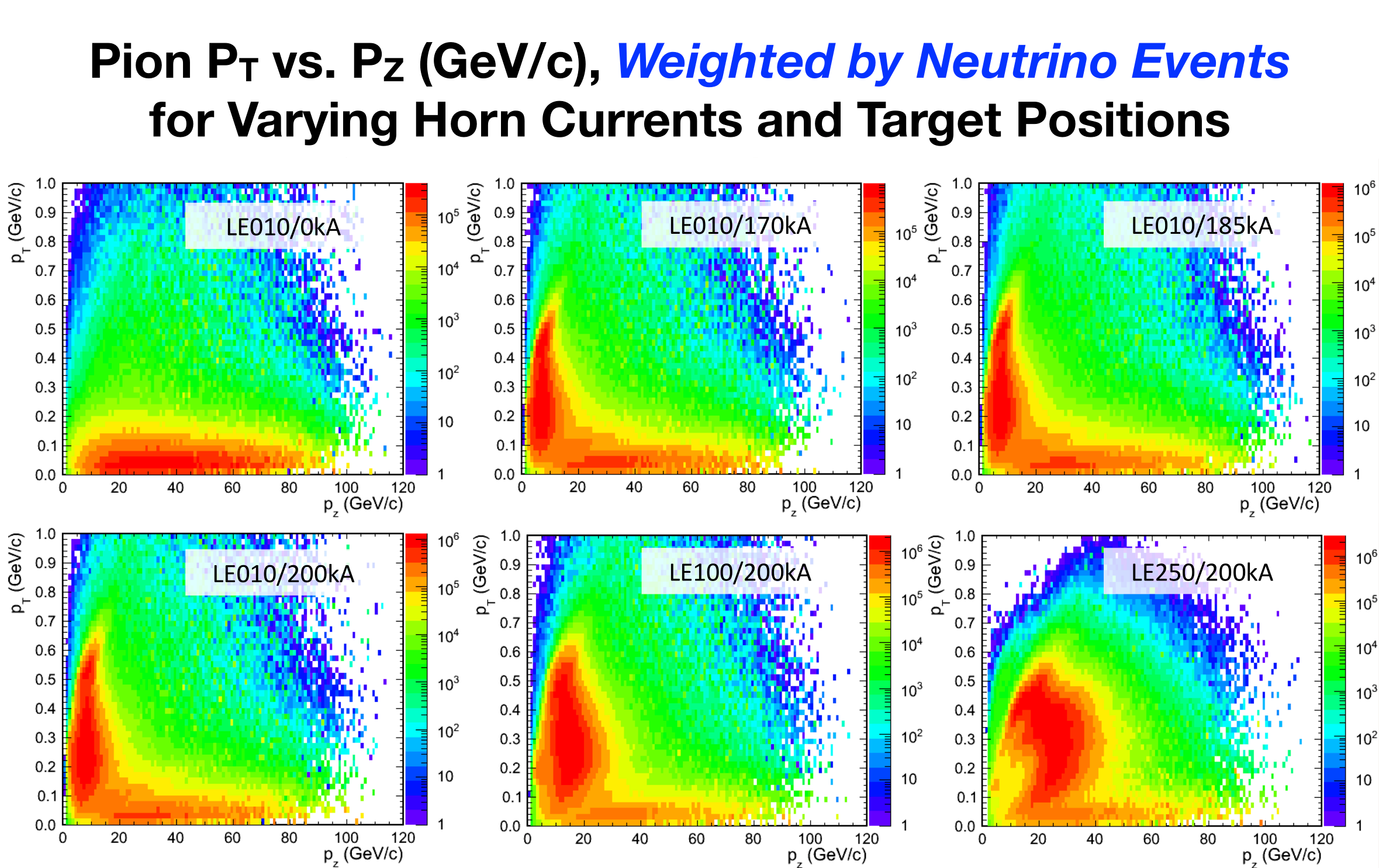,height=3.0in}
\caption{Shown here are pion $P_T$ vs. $P_Z$ distributions from hadrons produced on the primary target for a variety of horn currents and target positions. Events are weighted by neutrino events observed in the detector. 0kA, 170kA, etc. label the horn current. Higher currents focus higher $P_T$Õs. LE010, LE100, etc. label the target position. Moving the target back (LE100 $>$ LE010) focuses higher energy pions. 
\label{fig:specrunpi}}
\end{figure}

The largest uncertainty when estimating the neutrino flux is the hadron production spectra at the target. 
By utilizing the variable spectra at the NuMI beam-line, it is possible to fit for the various hadron production parameters in the MC. We do this by varying the focusing horn current (to focus pions of different $P_T$) and by varying the position of the target (to focus pions of different $x_F = P_Z / P_T$). 
See Fig. \ref{fig:specrunpi} for an illustration of the impact of changing the horn current amplitude and position of the target on the pion focusing performance. 

\section{Antineutrino Analysis}\label{sec:antinu}
The charged current (CC) signature is a muon from W exchange $\left( \bar{\nu} + p \rightarrow \mu^{+} + X\right)$. We examine only muon candidates originating in the fiducial tracker volume, and analyze momentum and sign in the MINOS Near Detector (we do not yet consider muons that stop inside $\mnv$). 

\subsection{Antineutrino Charged Current Quasi-elastic Events}
The charged current quasi-elastic (CCQE) channel $\left( \bar{\nu} + p \rightarrow \mu^{+} + n\right)$ is both clean and possesses a relatively large cross-section at the energies of the NuMI Low Energy (LE) configuration. The final state neutron is often invisible and the muon is relatively easy to identify and measure.

Presented here is a preliminary analysis conducted with $4 \times 10^{19}$ protons on target (POT) in the RHC LE beam configuration during detector construction, before starting our official physics run. The fiducial mass in the active tracker region used for this analysis was 2.86 tons of plastic scintillator.

\begin{figure}
\centering
\psfig{figure=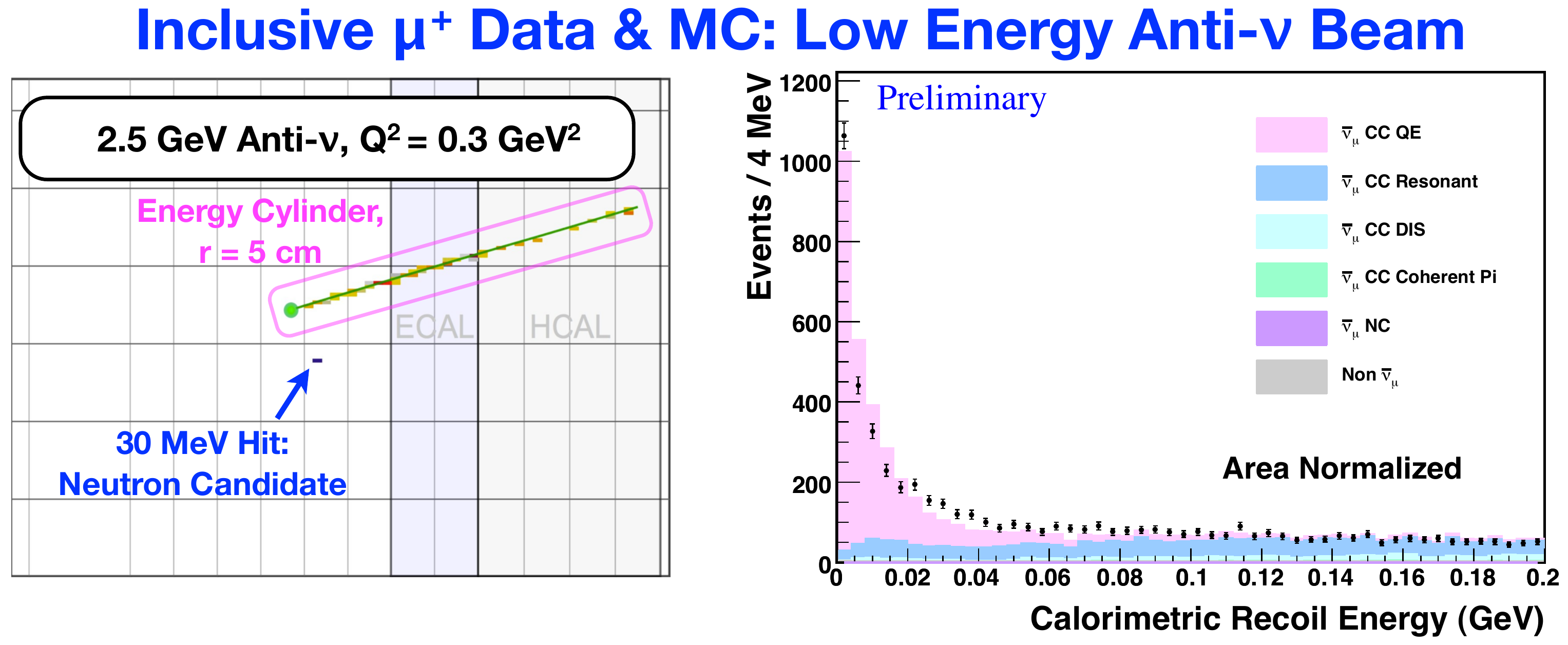,height=2.75in}
\caption{$\bar{\nu}$ charged-current ``extra-energy'' (recoil energy) data / simulation. 
\label{fig:nubarccrecoil}}
\end{figure}

The selection criterion is a $\mu^{+}$ originating in the fiducial $\mnv$ tracker volume well-reconstructed in the MINOS Near-Detector. In addition, very small ``recoil energy,'' or extra energy is required; where that energy was defined as all the energy outside a 5 cm radial cylinder around the track with a very tight (100 ns) time window. Figure \ref{fig:nubarccrecoil} shows a candidate event display and the extra energy distribution comparison between our data and MC (GENIE 2.6.2 with a GEANT4 detector simulation and custom optical model).

Under the QE hypothesis, we can reconstruct the neutrino energy and four-momentum transfer with only the muon information. Equation \ref{eq:erec} is for the neutrino energy; flip the nucleon masses for the antineutrino formula:
\begin{equation}
E_{\nu}^{rec} = \frac{ m_{p}^2 - \left( m_n - E_B \right)^2 - m_{\mu}^2 + 2 \left( m_n - E_B \right) E_{\mu} }{ 2 \left( m_n - E_B - E_{\mu} + p_{\mu} \cos \theta_{\mu} \right) }, 
\label{eq:erec}
\end{equation}
and with the neutrino energy in hand, we can calculate the four-momentum transfer, $Q^2$, using Eq. \ref{eq:qsqrd}:
\begin{equation}
Q_{rec}^2 = 2\,E_{\nu}^{rec}\,\left( E_{\mu} - p_{\mu} \cos \theta_{\mu} \right) - m_{\mu}^2 .
\label{eq:qsqrd}
\end{equation}

\begin{figure}
\centering
\psfig{figure=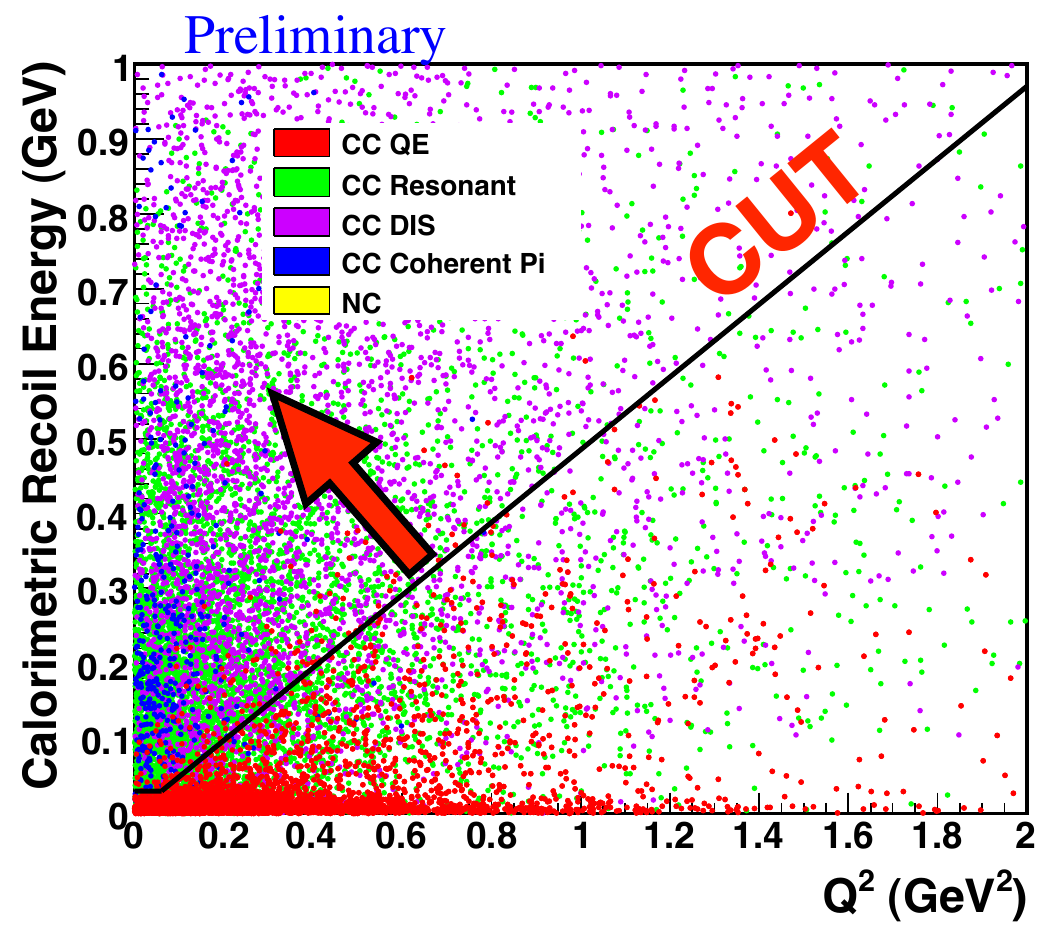,height=2.5in}
\caption{$\bar{\nu}$ charged-current ``extra-energy'' (recoil energy) versus $Q^2$ in simulation. 
\label{fig:nubarccrecoilcut}}
\end{figure}

By cutting on the extra-energy vs $Q^2$, we can produce a purified sample of CCQE candidates from our CC inclusive sample. Our cut is defined in Fig. \ref{fig:nubarccrecoilcut}. 

\begin{figure}
\centering
\psfig{figure=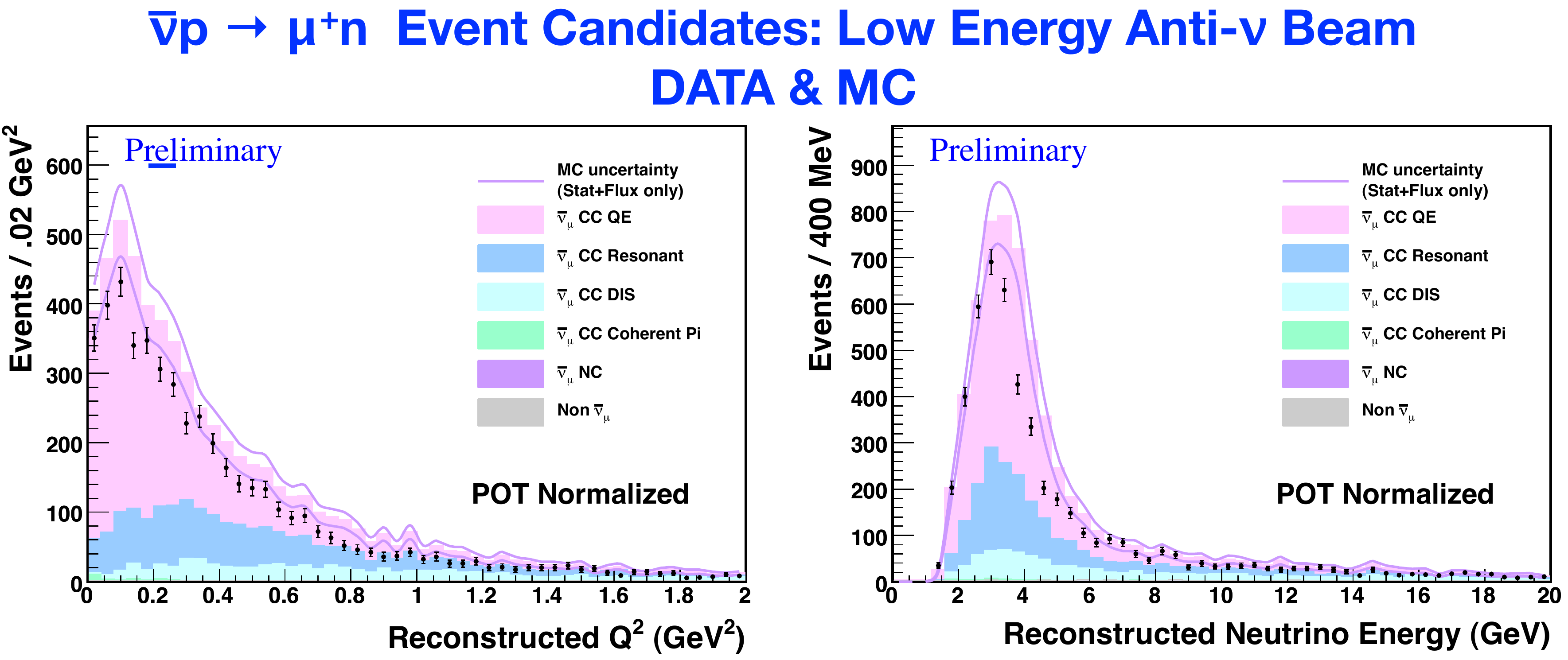,height=2.75in}
\caption{$\bar{\nu}$ data / simulation comparisons of energy and $Q^2$ reconstructed under the charged-current quasi-elastic hypothesis.
\label{fig:nubarccqeeq2}}
\end{figure}

Finally, in Fig. \ref{fig:nubarccqeeq2} we show the reconstructed neutrino energy and $Q^2$ compared to our current MC prediction, where absolute predictions are provided by our flux simulation. Note that the event deficit is flat in $Q^2$, but not neutrino energy.

\section{Calibration: The $\mnv$ Test Beam}\label{sec:tbeam}
In order to calibrate the detector response of $\mnv$, we conducted a Test Beam Experiment (TBE) at the FNAL Test Beam Facility (FTBF). The goal is to provide a hadronic response calibration, normalized to muon response, in a small-sized replica of the $\mnv$ detector.
The TBE detector was composed of roughly quarter-sized $\mnv$ planes that in every other aspect were as similar to the main $\mnv$ planes as possible. The TBE detector was read-out with photomultiplier tubes from the same set used to instrument the main detector, and all of the electronics and DAQ system were identical as well.

\begin{figure}
\centering
\psfig{figure=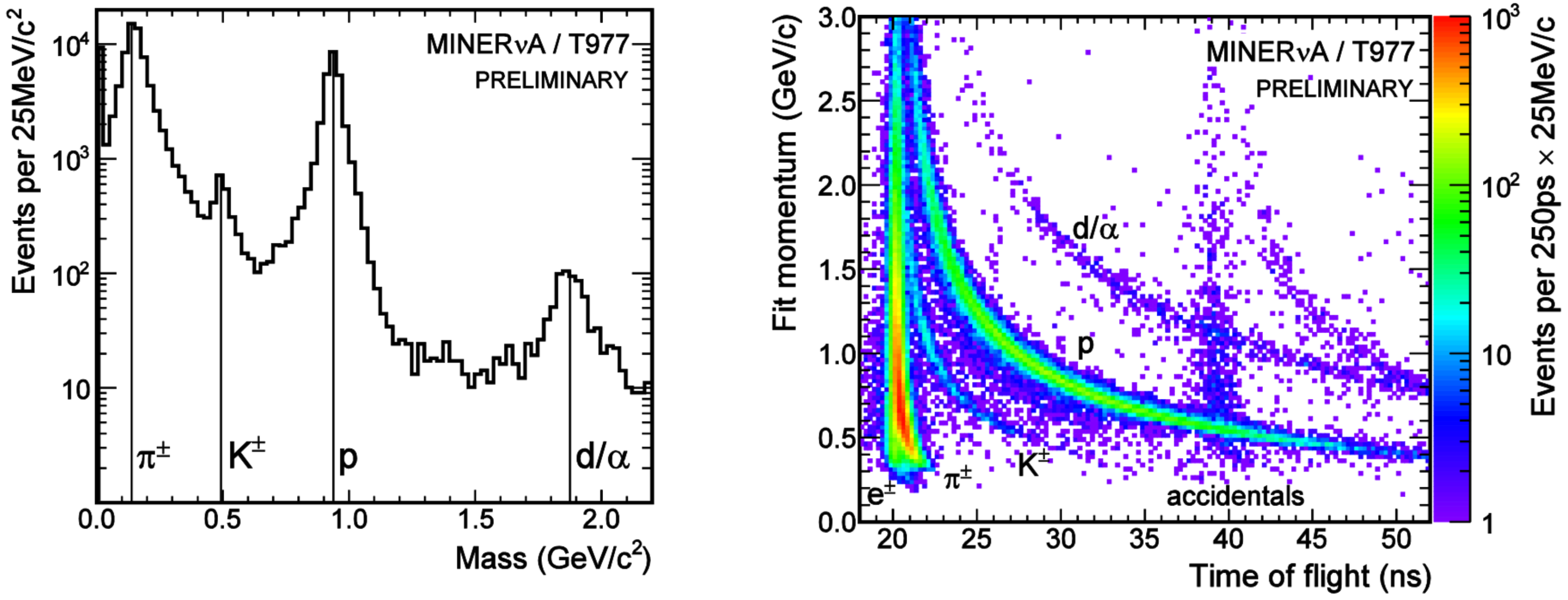,height=2.5in}
\caption{$\mnv$ Test Beam particle identification.
\label{fig:tbpid}}
\end{figure}

The TBE ran in two different detector configurations - one with 20 $\mnv$ Tracker planes and 20 ECAL planes, and another with 20 ECAL planes and 20 HCAL planes. We took roughly equal amounts of data in these two configurations, further dividing the data sets by magnet polarity (focusing either negative or positive pions in the tertiary beamline). Figure \ref{fig:tbpid} shows some particle ID distributions from the entire TBE data set. See Table \ref{tab:tbesamples} for a table of the data sample sizes.

\begin{table}[t]
\caption{$\mnv$ Sample sizes from the Test Beam Experiment \label{tab:tbesamples}}
\vspace{0.4cm}
\begin{center}
\begin{tabular}{|c|c|c|}
\hline
Configuration & Total Events & Passing All Beamline Selections \\
\hline
20 ECAL - 20 HCAL $\pi^{-}$ & 79,562 & 24,988 \\
\hline
20 ECAL - 20 HCAL $\pi^{+}$ & 77,639 & 32,935 \\
\hline
20 Tracker - 20 ECAL $\pi^{-}$ & 15,657 & 4,861 \\
\hline
20 Tracker - 20 ECAL $\pi^{+}$ & 93,667 & 43,587 \\
\hline
\hline
\end{tabular}
\end{center}
\end{table}

\section*{Acknowledgments}
This work was supported by DOE Grant No. DE-FG02-91ER40685.

\end{document}